# Simple theoretical model for parity-time-symmetric metasurfaces

**Jianlan Xie[1], Shaohua Dong[2], Bei Yan[1], Yuchen Peng[1], Jianjun Liu*,[1], Chengwei Qiu[2], Shuangchun Wen[1]**

[1]*Key Laboratory for Micro/Nano Optoelectronic Devices of Ministry of Education & Hunan Provincial Key Laboratory of Low-Dimensional Structural Physics and Devices, School of Physics and Electronics, Hunan University, Changsha 410082, China*

[2]*Department of Electrical and Computer Engineering, National University of Singapore, Engineering Drive, Singapore 117583, Singapore*

*E-mail: jianjun.liu@hnu.edu.cn

**Abstract:** Many new possibilities to observe and use novel physical effects are discovered at so called exceptional points (EPs). This is done by using parity-time (PT) -symmetric non-Hermitian systems and balancing gains and losses. When combined with EP-physics, recently, metasurfaces have shown greater abilities for wave manipulation than conventional metasurface systems. However, the solving process for EPs usually requires the transfer matrix method (TMM) or a parametric sweep, which are both complex and time-consuming. In this Letter, we develop a simple theoretical model, which is based on acoustic equivalent-circuit theory and can find the analytic solutions for EPs directly. As a proof of concept, PT-symmetric acoustic metasurfaces are studied to test the theoretical model, which enables unidirectional antireflection effects at EPs. In addition, finite element method (FEM) simulations are performed to study these EP solutions using the theoretical model for different mediums, wavelengths, angles of incidence, and gain-loss ratios. Our work offers a simple and powerful theoretical tool for designing PT-symmetric metasurfaces at EPs and may also be used for other classical wave systems.

**Keywords:** exceptional point, parity-time symmetry, non-Hermitian, acoustic metasurfaces, acoustic equivalent-circuit theory

# 1. Introduction

PT-symmetric non-Hermitian systems have revolutionized our understanding of condensed matter [1-3]. This is especially true for classical wave systems that can be fabricated and used for experiments [4-17]. In PT-symmetric systems of classical waves, loss is treated as a new degree of freedom that can be utilized effectively [4-26]. This can lead to many novel effects at EPs, which would be challenging to realize using conventional Hermitian systems [27-29]. On the other hand, metasurfaces provide a powerful platform to build these systems because they are freely designed structures especially in areas like optics and acoustics. For example, metasurface-based PT-symmetric systems [30-48] have shown many novel physical effects at EPs, such as unidirectional antireflection negative refraction [34,35], unidirectional antireflection subwavelength imaging [34-36], unidirectional invisibility [37,38], high-sensitivity sensing [39-41], optical force rectification [42], perfect invisibility [43], and asymmetric reflection or diffraction [46-48].

These novel physical effects depend on the realization of EPs, which are usually solved using the transfer matrix method (TMM) or parametric sweeps. However, the expected EPs can hardly be located if multiple solutions for the TMM exist. Additional measures are then needed to perform numerical simulations, which increase the complexity of the computation. For parametric sweeps, there is an intrinsic drawback, where some EP solutions could be missed due to the finite solution domain. In addition, both of these methods are time-consuming, especially when the system-parameters associated with the mediums (structure, incident wavelength, or angle) need to be adjusted. The required solution processes may then need to be repeated. Therefore, a simple theoretical model, which can directly and quickly locate EP solutions of PT-symmetric systems, could be a very valuable tool.

In this Letter, a simple theoretical model, which is based on acoustic equivalent-circuit theory, is proposed. It enables the direct identification of EP solutions of PT-symmetric systems. It is developed using two parallel acoustic metasurfaces (one is passive and another is active) that are kept at a constant distance from each other. In order to find EP solutions in this type of PT-symmetric system (i.e., ensuring that EPs exist), the admittances of the surrounding mediums of these two metasurfaces are under the intrinsic restrictions of Snell's law. In addition to this well-known prerequisite, another inherent prerequisite, which exists at the EP, can be analyzed using our theoretical model. Based on the theoretical model and the two inherent prerequisites, our PT-symmetric acoustic systems can realize the unidirectional antireflection negative refraction and positive refraction at EPs for arbitrary surrounding ordinary mediums and their different states of matter, arbitrary wavelengths, arbitrary incident angles, and arbitrary gain-loss ratios. Both theoretical and simulation results suggest that the model is accurate. Our model can be used to transfer difficult complex-parameter problems into easy-to-implement real-parameter problems. In addition, it can also be used to construct passive PT-symmetric metasurface systems.

## 2 Model and theory

### 2.1 A simple theoretical model for a CPA-amplifier system

PT-symmetric systems, which consist of two parallel metasurfaces [43], are in a PT-symmetric (PT-broken) state when the transmittance is $T < 1$ ($T > 1$). Furthermore, the two states and the associated eigenvalues coalesce at the phase-transition point (i.e., EP) when $T = 1$. According to the relationship between $T$, left reflectance $R_L$, and right reflectance $R_R$, the expression $\sqrt{R_L R_R} = |T - 1|$ yields $R_L = 0$ or $R_R = 0$ at the EP, which produces a unidirectional antireflection effect. In this Letter, we take $R_L = 0$ as an example – see Fig. 1(a). Total transmission is realized and the left reflectance is zero ($T = 1$, $R_L = 0$) at the EPs when the acoustic wave is incident from the left. However, there can be a very large right reflectance, when the wave is incident from the right (in this case), according to the relationship above despite $T = 1$. Therefore, to solve the EPs, it is sufficient to analyze the condition that corresponds to $T = 1$ using the TMM. However, it is a tedious process to solve EPs directly using the TMM without the help of other mathematical tools. In addition, this process often yields more than one solution, which can be divided (based on their physical mechanisms) into a coherent perfect absorber and amplifier (CPA-amplifier) system and an antireflection coating (ARCL-ARCR) system. However, the physical mechanisms of these two systems cannot be distinguished and accurately reflected by the TMM. Therefore, it would be helpful to develop a simple theoretical model for each system to distinguish and accurately reflect their different physical mechanisms. First, we take the CPA-amplifier system as an example – see Fig. 1.

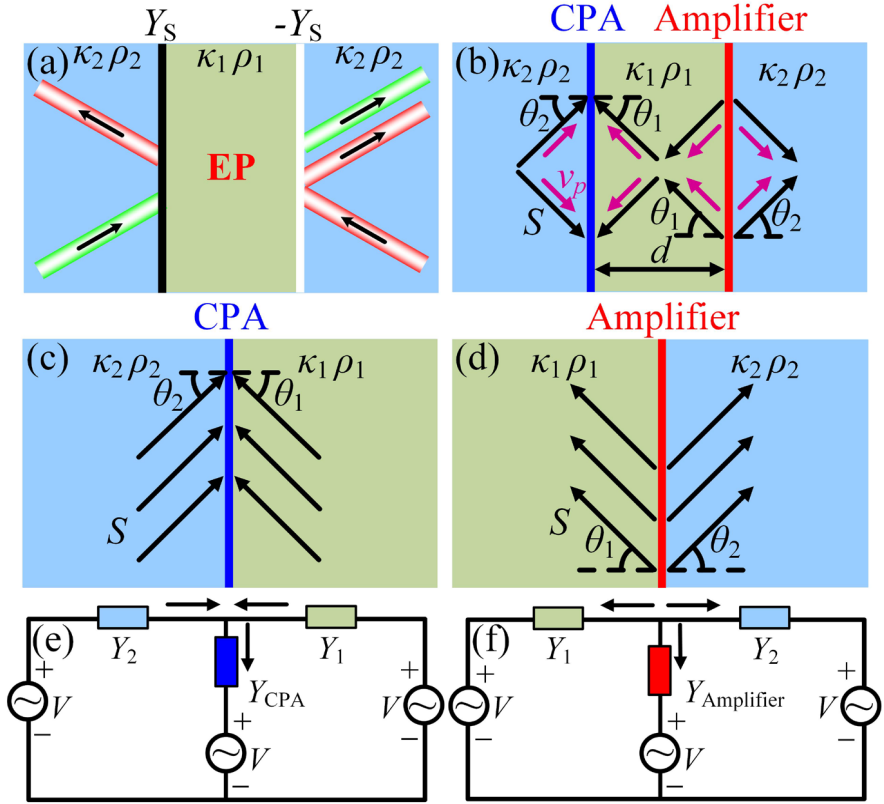

FIG.1. (a) Unidirectional antireflection effects at the EPs of PT-symmetric metasurfaces. (b) Unidirectional antireflection negative refraction of PT-symmetric metasurfaces. The light-green region indicates the middle medium, the light-blue region indicates the left/right medium, black and pink arrows indicate the Poynting vector direction (i.e., energy flow density direction) and the group-velocity direction of the acoustic waves, respectively. The separation of the CPA-amplifier system is also shown: (c) CPA metasurface, (d) amplifier metasurface. A simple theoretical model of a CPA-amplifier system, which consists of two acoustic equivalent-circuits: (e) Acoustic equivalent-circuit of CPA metasurface. (f) Acoustic equivalent-circuit of amplifier metasurface.

As shown in Fig. 1(b), according to previous studies [34,35,43,45], for the CPA-amplifier system, the system enables unidirectional antireflection negative refraction without using negative refractive materials and without considering impedance matching. Considering the mechanisms of two metasurfaces in the CPA-amplifier system, the directions of the Poynting vector and group-velocity direction in the system are shown in Fig. 1(b). The CPA metasurface, the acoustic waves, which are incident from its left and right sides, are absorbed perfectly. On the other hand, for the amplifier metasurface, the effect is inverse, i.e., it emits the acoustic waves to its left and right side. When the absorption and emission are balanced, the system realizes unidirectional antireflection negative refraction. Based on the mechanism of the CPA and amplifier metasurfaces, Fig. 1(b) can be divided into the two parts illustrated in Figs. 1(c) and 1(d).

When the thickness of the metasurface is much smaller than the wavelength of incidence, acoustic equivalent-circuit theory [49-51] can be used to equate the acoustic waves in Fig. 1(c) into the acoustic equivalent-circuit shown in Fig. 1(e). To analyze it more easily, the CPA metasurface and its surrounding mediums are represented by the admittance (the reciprocal of the impedance). Furthermore, according to Kirchhoff's current law, their admittances should satisfy the following relationship [52],

$$Y_{\mathrm{CPA}} = Y_1 \cos\theta_1 + Y_2 \cos\theta_2, \qquad (1)$$

where $Y_{\mathrm{CPA}}$ is the admittance of the CPA metasurface; $Y_i$, $\theta_i$, and $Y_i\cos\theta_i$ are the admittance, angle of incidence, and wave admittance of the surrounding mediums in the different regions, respectively ($i = 1, 2$, corresponds to the middle or left/right medium, respectively). Similarly, the acoustic waves in Fig. 1(d) can be equated using the acoustic equivalent-circuit shown in Fig. 1(f), where the acoustic waves are incident from the left and right of the amplifier metasurface. According to Kirchhoff's current law, their admittances should satisfy the following relationship [52]:

$$Y_{\mathrm{Amplifier}} = Y_1 \cos\theta_1 - Y_2 \cos\theta_2. \qquad (2)$$

Therefore, the CPA-amplifier system can be equated to a simple theoretical model, which consists of the two acoustic equivalent circuits shown in Figs. 1(e) and 1(f). In other words, a CPA-amplifier system can be realized when the PT-symmetric metasurfaces satisfy Eqs. (1) and (2).

**2.2 A simple theoretical model of an ARCL-ARCR system**

Based on the above derivation, a simple theoretical model of an ARCL-ARCR system can be obtained using the same method (see Part I of the Supplementary Materials). According to the model, the following relations can be obtained

$$Y_{ARCL} = -Y_1 \cos\theta_1 + Y_2 \cos\theta_2, \tag{3}$$

$$Y_{ARCR} = Y_1 \cos\theta_1 - Y_2 \cos\theta_2. \tag{4}$$

Both the theoretical models of the CPA-amplifier and the ARCL-ARCR systems mentioned above can be confirmed theoretically using a TMM (see Part II of the Supplementary Materials). They will also be verified by the simulation later in this Letter. Previous studies [34-36] treated the whole system as a two-port transmission-line network and solved EPs with the TMM. However, while they succeed in simplifying the system, these researches do not simplify the process to find solutions. Using the theoretical model, it can be seen that, in this Letter, we construct an acoustic equivalent-circuit for each metasurface based on acoustic equivalent-circuit theory [49-51]. This simplifies the process of finding solutions and highlights the corresponding physical mechanisms that have theoretical significance.

In acoustics, the property of the medium is often represented by the bulk modulus and density. For a surrounding ordinary medium of PT-symmetric metasurfaces, if its bulk modulus and density are $\kappa$ and $\rho$, respectively, its admittance $Y$ is defined as

$$Y = \sqrt{1/(\kappa\rho)}, \tag{5}$$

When the angle of incidence is $\theta$, the wave admittance $Y_W$ is defined as

$$Y_W = Y \cos\theta, \tag{6}$$

For a gain- or loss-medium, whose thickness is much smaller than the wavelength of incidence, provided its gain or loss are represented by the bulk modulus $\kappa_S$, $\kappa_S$ is related to its admittance $Y_S$ according to Ref. [43]

$$\kappa_S = i\omega_0 d_S/Y_S .\qquad(7)$$

Therefore, the thickness of metasurface is set to $d_S = 0.001\lambda$, where $\lambda$ is the wavelength of incidence, $\omega_0$ is the angular frequency in the background materials. Moreover, since the admittance of the metasurface is represented by the bulk modulus, its density $\rho_S$ is consistent with that of the background material $\rho_0$. In addition, according to Eqs. (1)-(4), the admittances of both metasurfaces and their surrounding ordinary mediums are independent of the distance between two metasurfaces. Hence, arbitrary lengths can be used, here, we set $d = 10$ mm.

## 3. Results and Discussion

To verify the accuracy of the theoretical model using simulations, the transmittance (using a logarithmic scale, i.e., Log$T$) of the system is calculated by sweeping the wave admittances of the surrounding mediums parametrically. For this, the admittances of PT-symmetric metasurfaces, the wavelength of incidence $\lambda$, and angle of incidence $\theta_2$ of this system are fixed (i.e., $\lambda = 5$mm, $\theta_2 = 30°$). The verification with regard to the uniformity of the numerical solutions (obtained from the parametric sweep) and analytical solutions (obtained from the theoretical model using Eqs. (1)-(4)) of EPs can be seen in Fig. 2.

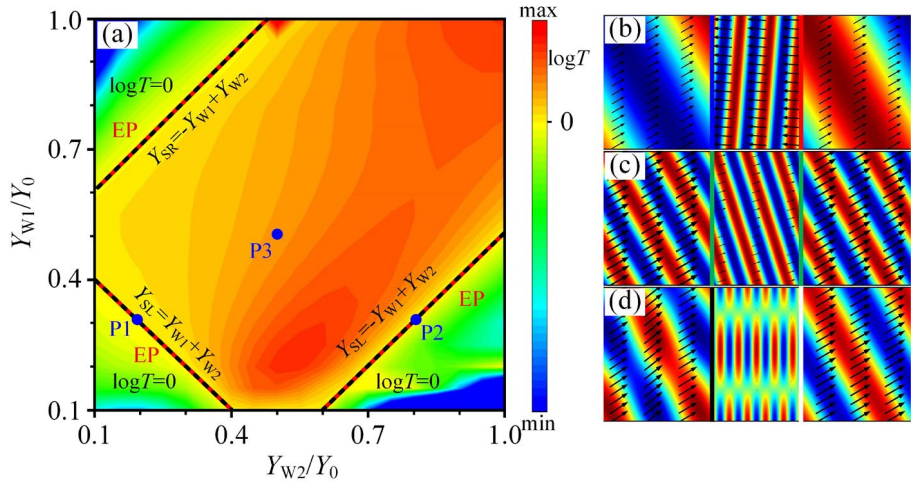

FIG. 2. (a) Log$T$ of the system calculated by sweeping $Y_{W1}/Y_0$ and $Y_{W2}/Y_0$ in the range of [0.1,1]. The admittances of left-side and right-side metasurfaces are fixed to $Y_{SL}= 0.5Y_0$ and $Y_{SR}= -0.5Y_0$. $Y_{SL}$ and $Y_{SR}$ can be fixed to arbitrary values, we take into account the sweep range of wave

admittances and combine Eqs. (1)-(4)). $Y_{W1}$ and $Y_{W2}$ denote the wave admittances of the middle and left/right mediums, respectively. $Y_0$ denotes the admittance of the background material, which is often set to air for gaseous systems. The red dashed lines indicate all numerical solutions of EPs in the range for the above conditions, which correspond to Log$T$ = 0. The black dashed lines indicate the analytic solutions of EPs obtained using the theoretical model, which correspond to equations in this figure. (b) Total sound pressure field of the CPA-amplifier system that corresponds to P1 in (a), where $Y_{W1}/Y_0$ = 0.3 and $Y_{W2}/Y_0$ = 0.2. (c) The total sound pressure field of ARCL-ARCR system that corresponds to P2 in (a), where $Y_{W1}/Y_0$ = 0.3 and $Y_{W2}/Y_0$ = 0.8. (d) Total sound pressure field of the ordinary point (i.e., $T \neq 1$) that corresponds to P3 in (a), where $Y_{W1}/Y_0$ = 0.5 and $Y_{W2}/Y_0$ = 0.5.

Figure 2(a) shows that the EPs can be located using a parametric sweep for the above conditions, as the red dashed lines indicate, which correspond to Log$T$ = 0. Moreover, the different unidirectional antireflection effects can be distinguished using further simulations, as shown in Figs. 2(b) and 2(c). However, these effects cannot be realized when the parameters do not belong to red dashed lines, as shown in Fig. 2(d). This process is quite involved and requires additional simulations to distinguish different systems. The analytical solutions of EPs, which are obtained with the theoretical model, not only can correspond to the numerical solutions of EPs one by one but also can directly distinguish different systems without additional simulations. This can be seen from the equations in Fig. 2(a). The black dashed lines in Fig 2(a) represent the analytical solutions of the EPs obtained with the theoretical model. The black dashed line in the lower-left corner represents the equation for the CPA metasurface, which is $Y_{SL} = Y_{W1} + Y_{W2}$ and corresponds to Eq. (1) when combined with Eq. (6). The black dashed line in the lower-right corner represents the equation for the ARCR metasurface, which is $Y_{SL} = -Y_{W1} + Y_{W2}$ and corresponds to Eq. (3) when combined with Eq. (6). In addition, the red dashed line in the upper-left corner of Fig. 2(a) represents also the solution for the EPs, which enables total transmission. However, $R_L$ is not zero at this time because only the case of waves incident from the left is considered here. According to the theory discussed in Section 2.1, $R_R$ is zero at this time, which corresponds to EPs when waves are incident from the right. The black dash line in the upper-left corner represents the equation for the ARCR metasurface when the wave is incident from the right, which is $Y_{SR} = -Y_{W1} + Y_{W2}$ and corresponds to Eq. (3) when combined with Eq. (6). This further illustrates the accuracy of the theoretical model.

The first inherent prerequisite for the EP is usually that a total reflection cannot be achieved, i.e., $\sqrt{\kappa_1\rho_2/\kappa_2\rho_1} \cdot \sin\theta_2 < 1$. This condition can be deduced from Snell's law of refraction. The second inherent prerequisite (an EP exists) can be analyzed using our theoretical model. In other words, the difference between the wave admittances of the middle and left/right mediums cannot exceed one order of magnitude, i.e., $0.1 < \sqrt{\kappa_1\rho_1\cos^2\theta_2/\kappa_2\rho_2\cos^2\theta_1} < 10$. These two inherent prerequisites both can be verified with the simulation (see Part III of the Supplementary Materials). Based on the theoretical model for the two inherent prerequisites, the parameters of the PT-symmetric metasurfaces, which are required for the solution of EPs, can be obtained when its surrounding medium, incident wavelength, and angle of incidence are determined (see Parts IV and V of the Supplementary Materials).

For an additional confirmation of the general applicability of the theoretical model, Log$T$ of the system is calculated for arbitrary conditions. By considering the gaseous system and setting $\kappa_i = b_i\kappa_0$ and $\rho_i = c_i\rho_0$ (where $\kappa_i$ and $\rho_i$ are the bulk modulus and density of the surrounding mediums, $\kappa_0$ and $\rho_0$ are the bulk modulus and density of air ($\kappa_0$ = 1.42×10$^5$ Pa, $\rho_0$ = 1.2 kg/m$^3$), $b_i$ and $c_i$ are arbitrary constants ($i$ = 1, 2, corresponds to middle or left/right medium, respectively)). By assuming constant $b_2$, $c_2$, $\theta_2$ and $\lambda$, the Log$T$ of the system is calculated by sweeping $b_1$ and $c_1$ in the range of [1,10]. The results are shown in Fig. 3.

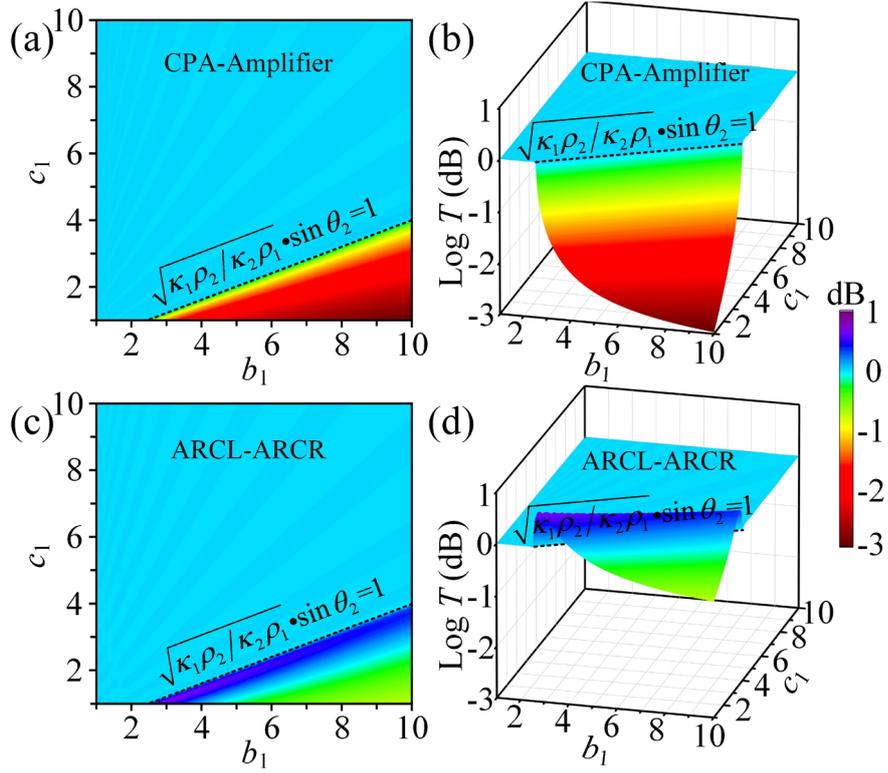

FIG. 3. Log$T$ of the system, which is calculated by sweeping $b_1$ and $c_1$ for the range [1,10], where the parameters of the metasurface are obtained using the theoretical model and $\lambda$=5mm, $b_2$=1.47, $c_2$=2.34, $\theta_2$=30°. (a) CPA-amplifier system. (b) Three-dimensional view of (a). (c) ARCL-ARCR system. (d) Three-dimensional view of (c). The dotted line in all figures represents the first inherent prerequisite ($\sqrt{\kappa_1\rho_2/\kappa_2\rho_1}\cdot\sin\theta_2=1$).

As can be seen from Fig. 3, total transmission (i.e., Log$T$ = 0) is realized in most regions when $b_1, c_1 \in [1,10]$, which corresponds to EPs. However, there exists a small part of the region, where the system cannot realize total transmission. The reason for this is that surrounding mediums in these regions cannot satisfy the first inherent prerequisite. Therefore, based on the theoretical model, the CPA-amplifier and ARCL-ARCR systems can be realized using the PT-symmetric metasurfaces for arbitrary middle mediums, provided the two inherent prerequisites are satisfied. In addition, by analyzing the left/right mediums in the same manner (see Part VI of the Supplementary Materials), it also can be found that the theoretical model can be applied to arbitrary left/right mediums. For more confirmations that the theoretical model is also universally applicable to arbitrary incident wavelengths and angles, the wavelengths and angles of incidence are analyzed in the same way (see Part VII of the Supplementary Materials).

The above results demonstrate the universality of the theoretical model in the gain-loss balanced system. For a more general gain-loss imbalanced system, a corresponding simple theoretical model can also be obtained. It can also verify that this theoretical model is universally applicable to PT-symmetric metasurfaces that satisfy arbitrary gain-loss ratios (see Part VIII of the Supplementary Materials), which is important for practical applications. Furthermore, the required parameters of surrounding ordinary mediums for EPs can be obtained using the theoretical model, when the parameters of metasurfaces are fixed (see Part IX of the Supplementary Materials). This makes it possible to convert the difficult complex-parameter problem into an easy-to-implement real-parameter problem and greatly reduces the difficulty to realize EPs in a real-life system. In addition, our theoretical model can also be applied to other states of matter (see Part X of the Supplementary Materials). Finally, our theoretical model is an excellent guide to construct PT-symmetric metasurfaces for a real-life system, and the ARCL-ARCR system (with only passive metasurfaces) can be constructed using the theoretical model, which greatly reduces the difficulty to obtain a real-life ARCL-ARCR system (see Part XI of the Supplementary Materials).

For PT-symmetric metasurfaces in other classical wave fields, provided the novel effects at the EPs correspond to either a CPA-amplifier or ARCL-ARCR system, these effects can be analyzed using the same method. Hence, an equivalent simple theoretical model can be constructed for these systems. In addition, for other systems with non-Hermitian metasurfaces, the method to solve the EPs is the parametric sweep. Most of them are swept directly using relevant parameters like the refractive index, which often cannot produce the analytical solutions for EPs. However, in this Letter, we propose a simple theoretical model of the unidirectional antireflection effects at EPs and obtain the analytical solutions for the EPs. The solutions are only determined by the wave admittances of surrounding mediums, which provides a new idea for solutions of EPs in these systems. Therefore, our theoretical model is also applicable to PT-symmetric metasurfaces in other classical wave systems and can provide a valuable theoretical guide to find solutions of EPs in non-Hermitian metasurface systems.

## 4. Conclusions

In this Letter, a simple theoretical model was proposed, which was based on the physical mechanisms at EPs of PT-symmetric acoustic metasurfaces. We used acoustic equivalent-circuit theory and verified the accuracy of the model. The analytical solutions, which were obtained using the theoretical model, could solve the EPs directly, with high accuracy and comprehensively. Based on the theoretical model for two inherent prerequisites, EPs of PT-symmetric acoustic systems were obtained for arbitrary conditions. Our theoretical model helps facilitate the development of real-life acoustic PT-symmetric systems. In addition, the model can also be used for other classical wave systems, and it offers theoretical improvements for solving the EPs of non-Hermitian metasurface systems.

**Acknowledgments**

This work was supported by the National Natural Science Foundation of China (Grants No. 61405058, No. 62075059, and No. 12004258), the Natural Science Foundation of Hunan Province (Grants No. 2017JJ2048 and No. 2020JJ4161), and the Fundamental Research Funds for the Central Universities (Grant No. 531118040112). The authors acknowledge Professor J. Q. Liu for software sponsorship.

**Reference**

[1] C. M. Bender and S. Boettcher, "Real spectra in non-Hermitian Hamiltonians having PT symmetry," Phys. Rev. Lett. 80(24): 5243-5246 (1998).

[2] C. M. Bender, D. C. Brody, and H. F. Jones, "Complex extension of quantum mechanics," Phys. Rev. Lett. 89(27): 270401 (2002).

[3] C. M. Bender, "Making sense of non-Hermitian Hamiltonians," Rep. Prog. Phys. 70(6): 947 (2007).

[4] K. G. Makris, R. El-Ganainy, D. N. Christodoulides, and Z. H. Musslimani,

"Beam dynamics in PT symmetric optical lattices," Phys. Rev. Lett. 100(10): 103904 (2008).

[5] A. Guo, G. J. Salamo, D. Duchesne, R. Morandotti, M. Volatier-Ravat, V. Aimez, G. A. Siviloglou, and D. N. Christodoulides, "Observation of PT-symmetry breaking in complex optical potentials," Phys. Rev. Lett. 103(9): 093902 (2009).

[6] C. E. Rüter, K. G. Makris, R. El-Ganainy, D. N. Christodoulides, M. Segev and D. Kip, "Observation of parity-time symmetry in optics," Nat. Phys. 6(3): 192-195 (2010).

[7] A. Regensburger, C. Bersch, M. -A. Miri, G. Onishchukov, D. N. Christodoulides, and U. Peschel, "Parity-time synthetic photonic lattices," Nature 488(7410): 167-171 (2012).

[8] H. Hodaei, M.-A. Miri, M. Heinrich, D. N. Christodoulides, and M. Khajavikhan, "Parity-time-symmetric microring lasers," Science 346(6212): 975 (2014).

[9] R. Fleury, D. Sounas, and A. Alù, "An invisible acoustic sensor based on parity-time symmetry," Nat. Commun. 6(1): 5905 (2015).

[10] C. Shi, M. Dubois, Y. Chen, L. Cheng, H. Ramezani, Y. Wang, and X. Zhang, "Accessing the exceptional points of parity-time symmetric acoustics," Nat. Commun. 7(1): 11110 (2016).

[11] W. Chen, Ş. K. Özdemir, G. Zhao, J. Wiersig, and L. Yang, "Exceptional points enhance sensing in an optical microcavity," Nature 548(7666): 192-196 (2017).

[12] T. Liu, X. Zhu, F. Chen, S. Liang, and J. Zhu, "Unidirectional wave vector manipulation in two-dimensional space with an all passive acoustic parity-time-symmetric metamaterials crystal," Phys. Rev. Lett. 120(12): 124502 (2018).

[13] Y. Li, Y. -G. Peng, L. Han, M. -A. Miri, W. Li, M. Xiao, X. -F. Zhu, J. Zhao, A. Alù, S. Fan, and C. -W. Qiu, "Anti-parity-time symmetry in diffusive systems," Science 346(6436): 170-173 (2019).

[14] A. Li, J. Dong, J. Wang, Z. Cheng, J. S. Ho, D. Zhang, J. Wen, X. -L. Zhang, C. T. Chan, A. Alù, C. -W. Qiu, and L. Chen, "Hamiltonian hopping for efficient chiral mode switching in encircling exceptional points," Phys. Rev. Lett. 125(18):

187403 (2020)

[15] Y. Hadad and N. Engheta, "Possibility for inhibited spontaneous emission in electromagnetically open parity-time-symmetric guiding structures," P. Natl. Acad. Sci. USA 117(11): 5576-5581 (2020).

[16] H. -Z. Chen, T. Liu, H. -Y. Luan, R. -J. Liu, X. -Y. Wang, X. -F. Zhu, Y. -B. Li, Z. -M. Gu, S. -J. Liang, H. Gao, L. Lu, L. Ge, S. Zhang, J. Zhu, and R. M. Ma, "Revealing the missing dimension at an exceptional point," Nat. Phys. 16(5): 571-578 (2020).

[17] S. Weidemann, M. Kremer, T. Helbig, T. Hofmann, A. Stegmaier, M. Greiter, R. Thomale, and A. Szameit, "Topological funneling of light," Science 368(6488): 311-314 (2020).

[18] H. Li, A. Mekawy, A. Krasnok, and A. Alù, "Virtual parity-time symmetry," Phys. Rev. Lett. 124(19): 193901 (2020).

[19] A. Krasnok and A. Alù, "Active nanophotonics," P. IEEE 108(5): 628-654 (2020).

[20] S. Liu, S. Ma, C. Yang, L. Zhang, W. Gao, Y. J. Xiang, T. J. Cui, and S. Zhang, "Gain- and loss-induced topological insulating phase in a non-Hermitian electrical circuit," Phys. Rev. Applied 13(1): 014047 (2020).

[21] H. Zhang, R. Huang, S. -D. Zhang, Y. Li, C. -W. Qiu, F. Nori, and H. Jing, "Breaking anti-PT symmetry by spinning a resonator," Nano Lett. 20(10): 7594-7599 (2020).

[22] Y. Li, W. Li, and C. -W. Qiu, "Can scaling analysis be used to interpret the anti-parity-time symmetry in heat transfer?" arXiv:2008.10990.

[23] A. Bergman, R. Duggan, K. Sharma, M. Tur, A. Zadok, and A. Alù, "Observation of anti-parity-time-symmetry, phase transitions and exceptional points in an optical fibre," Nat. Commun. 12(1): 486 (2021).

[24] Z. Dong, H. -J. Kim, H. Cui, C. Li, C. -W. Qiu, and J. S. Ho, "Wireless magnetic actuation with a bistable parity-time-symmetric circuit," Phys. Rev. Applied 15(02): 024023 (2021).

[25] S. Xia, D. Kaltsas, D. Song, L. Komis, J. Xu, A. Szameit, H. Buljan, K. G. Makris, and Z. Chen, "Nonlinear tuning of PT symmetry and non-Hermitian

topological state," Science 372(6537): 72-76 (2021).

[26] A. Krasnok, N. Nefedkin, and A. Alù, "Parity-time symmetry and exceptional points: a tutorial," arXiv:2103.08135.

[27] D. Christodoulides and J. Yang, Parity-time symmetry and its applications (Springer, Berlin, 2018).

[28] M.-A. Miri and A. Alù, "Exceptional points in optics and photonics," Science 363(6422): eaar7709 (2019).

[29] Ş. K. Özdemir, S. Rotter, F. Nori, and L. Yang, "Parity-time symmetry and exceptional points in photonics," Nat. Mater. 18(8): 783 (2019).

[30] Y. D. Chong, L. Ge, and A. D. Stone, "PT-Symmetry breaking and laser-absorber modes in optical scattering systems," Phys. Rev. Lett. 106(9): 093902 (2011).

[31] L. Ge, Y. D. Chong, and A. D. Stone, "Conservation relations and anisotropic transmission resonances in one-dimensional PT-symmetric photonic heterostructures," Phys. Rev. A 85(2): 023802 (2012).

[32] J. D. H. Rivero and L. Ge, "Time-reversal-invariant scaling of light propagation in one-dimensional non-Hermitian systems," Phys. Rev. A 100(2): 023819 (2019).

[33] H. Wu, X. Yang, D. Deng, and H. Liu, "Reflectionless phenomenon in PT-symmetric periodic structures of one-dimensional two-material optical waveguide networks," Phys. Rev. A 100(3): 033832 (2019).

[34] R. Fleury, D. L. Sounas, and A. Alù, "Negative refraction and planar focusing based on parity-time symmetric metasurfaces," Phys. Rev. Lett. 113(2): 023903 (2014).

[35] F. Monticone, C. A. Valagiannopoulos, and A. Alù, "Parity-Time symmetric nonlocal metasurfaces: all-angle negative refraction and volumetric imaging," Phys. Rev. X 6(4): 041018 (2016).

[36] C. A. Valagiannopoulos, F. Monticone, and A. Alù, "PT-symmetric planar devices for field transformation and imaging," J. Optics 18(4): 044028 (2016).

[37] D. L. Sounas, R. Fleury, and A. Alù, "Unidirectional cloaking based on metasurfaces with balanced loss and gain," Phys. Rev. Applied. 4(1): 014005

(2015).

[38] H. Li, M. Rosendo-López, Y. Zhu, X. Fan, D. Torrent, B. Liang, J. Cheng, and J. Christensen, "Ultrathin acoustic parity-time symmetric metasurface cloak," Research 2019: 8345683 (2019).

[39] S. Xiao, J. Gear, S. Rotter, and J. Li, "Effective PT-symmetric metasurfaces for subwavelength amplified sensing," New J. Phys. 18(8): 085004 (2016).

[40] P. -Y. Chen and J. Jung, "PT symmetry and singularity-enhanced sensing based on photoexcited graphene metasurfaces," Phys. Rev. Applied 5(6): 064018 (2016).

[41] Y. J. Zhang, H. Kwon, M. -A. M., E. Kallos, H. Cano-Garcia, M. S. Tong, and A. Alù, "Noninvasive glucose sensor based on parity-time symmetry," Phys. Rev. Applied 11(4): 044049 (2019).

[42] R. Alaee, B. Gurlek, J. Christensen, M. Kadic, "Optical force rectifiers based on PT-symmetric metasurfaces," Phys. Rev. B 97(19): 195420 (2018).

[43] J. Luo, J. Li, and Y. Lai, "Electromagnetic impurity-immunity induced by parity-time symmetry," Phys. Rev. X 8(3): 031035 (2018).

[44] M. Sakhdari, N. M. Estakhri, H. Bagci, and P. -Y. Chen, "Low-threshold lasing and coherent perfect absorption in generalized PT-symmetric optical structures," Phys. Rev. Applied 10(2): 024030 (2018).

[45] J. Lan, X. Zhang, L. Wang, Y. Lai, and X. Liu, "Bidirectional acoustic negative refraction based on a pair of metasurfaces with both local and global PT-symmetries," Sci. Rep.-UK 10(1):10794 (2020).

[46] X. Wang, X. Fang, D. Mao, Y. Jing, and Y. Li, "Extremely asymmetrical acoustic metasurface mirror at the exceptional point," Phys. Rev. Lett. 123(21): 214302 (2019).

[47] S. Dong, G. Hu, Q. Wang, Y. Jia, Q. Zhang, G. Cao, J. Wang, S. Chen, D. Fan, W. Jiang, Y. Li, A. Alù, and C.-W. Qiu, "Loss-Assisted Metasurface at an Exceptional Point." ACS Photon. 7(12): 3321 (2020).

[48] Y. Yang, H. Jia, Y. Bi, H. Zhao, and J. Yang, "Experimental demonstration of an acoustic asymmetric diffraction grating based on passive parity-time-symmetric

medium," Phys. Rev. Applied 12(3): 034040 (2019).

[49] F. Bongard, H. Lissek, and J. R. Mosig, "Acoustic transmission line metamaterial with negative/zero/positive refractive index," Phys. Rev. B 82(9): 094306 (2010).

[50] C. Shen, J. Xu, N. X. Fang, and Y. Jing, "Anisotropic complementary acoustic metamaterial for canceling out aberrating layers," Phys. Rev. X 4(4): 041033 (2014).

[51] J. Zhang, Y. Cheng, and X. Liu, "Extraordinary acoustic transmission at low frequency by a tunable acoustic impedance metasurface based on coupled Mie resonators," Appl. Phys. Lett. 110(23): 233502 (2017).

[52] C. K. Alexander and M. Sadiku, Fundamentals of electric circuits. 6th ed. (McGraw-Hill Education, 2016).

# Supplementary Materials for "Simple theoretical model for parity-time-symmetric metasurfaces"

Jianlan Xie[1], Shaohua Dong[2], Bei Yan[1], Yuchen Peng[1], Jianjun Liu*[,1], Chengwei Qiu[2], Shuangchun Wen[1]

[1]*Key Laboratory for Micro/Nano Optoelectronic Devices of Ministry of Education & Hunan Provincial Key Laboratory of Low-Dimensional Structural Physics and Devices, School of Physics and Electronics, Hunan University, Changsha 410082, China*

[2]*Department of Electrical and Computer Engineering, National University of Singapore, Engineering Drive, Singapore 117583, Singapore*

## I. The simple theoretical model of an ARCL-ARCR system

In Part 2.1 of the main text, a simple theoretical model of the CPA-amplifier system was developed by analyzing the corresponding physical mechanisms of the CPA-amplifier system, using acoustic equivalent-circuit theory. with the help of the same method to analyze the ARCL-ARCR system, a corresponding simple theoretical model can be built – see Fig. S1.

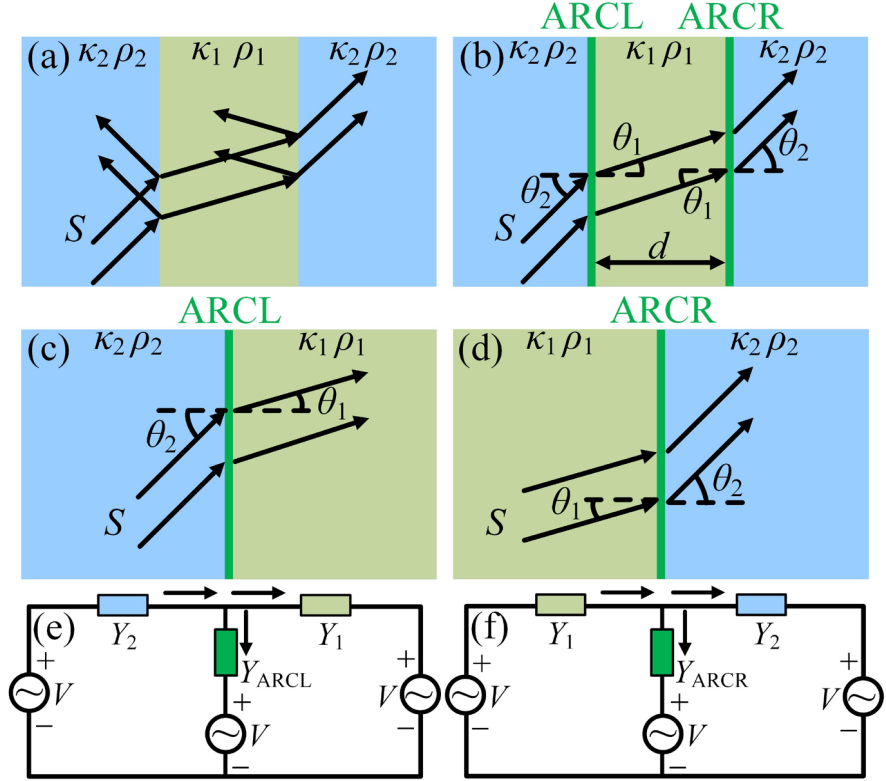

FIG. S1. (a) Ordinary positive refraction. (b) Unidirectional antireflection positive refraction of PT-symmetric acoustic metasurfaces, where the light-green region indicates the central medium, while he light-blue region indicates the left/right medium. The black arrows indicate the Poynting-vector direction of the acoustic waves. (c) ARCL metasurface. (d) ARCR metasurface. A simple theoretical model of an ARCL-ARCR system, which consists of two acoustic equivalent-circuits: (e) Acoustic equivalent-circuit of the ARCL metasurface. (f) Acoustic equivalent-circuit of the ARCR metasurface.

As shown in Fig. S1(a), for ordinary positive refraction, if the impedances of the middle and left/right mediums do not match, and the angle of incidence also does not satisfy the Brewster angle, there reflections can occur. Although the ultra-transparent effect can be realized by metamaterials [SR1] or bandgap structures [SR2-SR4], the implementation methods tend to be too complicated (and limited to a single frequency or angle). However, the ARCL-ARCR system of PT-symmetric metasurfaces can realize unidirectional antireflection positive refraction and is neither limited by angle nor frequency – see Fig. S1(b).

Based on the mechanisms of the acoustic wave's Poynting vector in the

ARCL-ARCR system, Fig. S1(b) also can be divided into two parts - see Figs. S1(c) and S1(d). Using the acoustic equivalent-circuit theory [SR5-SR7], when the thickness of the metasurfaces is much smaller than the wavelength of incidence, the acoustic waves in Figs. S1(c) and S1(d) can be equated to acoustic equivalent-circuits in Figs. S1(e) and S1(f), respectively. Using Kirchhoff's current law [SR8], the relations that need to be satisfied for the ARCL and ARCR metasurfaces can be obtained are

$$Y_{ARCL} = -Y_1 \cos\theta_1 + Y_2 \cos\theta_2, \qquad (S1)$$

$$Y_{ARCR} = Y_1 \cos\theta_1 - Y_2 \cos\theta_2. \qquad (S2)$$

Here, $Y_{ARCL}$ and $Y_{ARCR}$ are the admittances of the ARCL and ARCR metasurfaces, while $Y_i$, $\theta_i$, and $Y_i\cos\theta_i$ are the admittances, angles of incidence, and wave admittances of surrounding mediums in different regions, respectively ($i$ = 1, 2, corresponds to the middle or left/right mediums, respectively). Therefore, when the PT-symmetric metasurfaces satisfy Eqs. (S1) and (S2) (which correspond to Eqs. (3) and (4) in the main text, respectively), an ARCL-ARCR system can be realized.

**II. Theoretical verification of the simple theoretical model**

In order to verify the accuracy of the theoretical model theoretically, the EPs of the PT-symmetric acoustic metasurfaces are solved using the transfer matrix method (TMM). According to previous studies [SR9], this requires that the gain and loss of the two metasurfaces are balanced for PT-symmetric systems with two parallel metasurfaces. In other words, the admittances of two metasurfaces are opposite to each other. Because only the transfer matrix of a system needs to be considered, both the CPA-amplifier and ARCL-ARCR systems can now be represented by the same model. Considering the cases that the acoustic waves are transverse, and the definition of acoustic admittance corresponds to the TE mode in optics [SR10,SR11], the CPA-amplifier (see Fig. 1(b) in the main text) and ARCL-ARCR systems (see Fig. S1(b)) can be equated to a parallel two-port transmission-line network model – see Fig. S2.

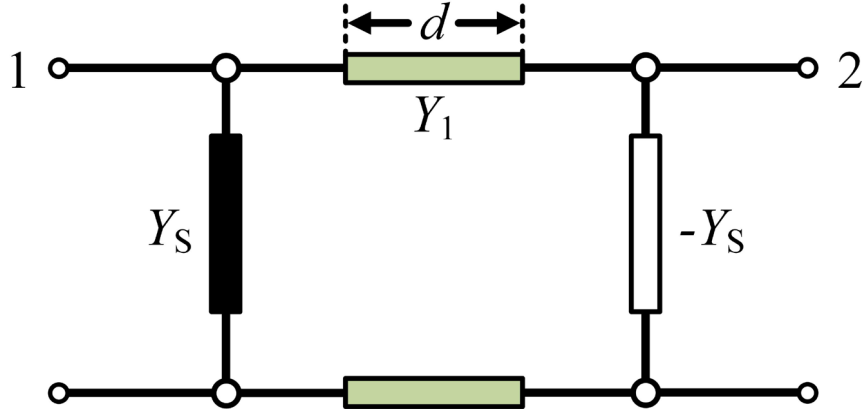

FIG. S2. The equivalent parallel two-port transmission-line network model.

According to TMM, if the middle medium between two metasurfaces is an ordinary medium, using the theory for a parallel two-port network model [SR12], the transfer matrix of the whole system can be expressed as

$$M = \begin{bmatrix} 1 & 0 \\ Y_S & 1 \end{bmatrix} \begin{bmatrix} \cos\varphi & -i\dfrac{1}{Y_1 \cos\theta_1}\sin\varphi \\ -iY_1 \cos\theta_1 \sin\varphi & \cos\varphi \end{bmatrix} \begin{bmatrix} 1 & 0 \\ -Y_S & 1 \end{bmatrix}. \quad (S3)$$

Here, $\pm Y_S$ is the admittance of the left-side metasurface and the right-side metasurface, respectively. $Y_1$, $\theta_1$, $\kappa_1$, and $\rho_1$ denote the admittance, angle of incidence, bulk modulus, and density of the middle medium, respectively. Furthermore, $\varphi$ is the phase difference, which is generated by the acoustic waves passing through the middle medium. By calculating the matrix, we obtain:

$$M = \begin{bmatrix} \dfrac{iY_S}{\sqrt{\dfrac{1}{\rho_1\kappa_1}}\cos\theta_1}\sin\varphi + \cos\varphi & -\dfrac{i}{\sqrt{\dfrac{1}{\rho_1\kappa_1}}\cos\theta_1}\sin\varphi \\ -i\sqrt{\dfrac{1}{\rho_1\kappa_1}}\cos\theta_1 \sin\varphi + \dfrac{i(Y_S)^2}{\sqrt{\dfrac{1}{\rho_1\kappa_1}}\cos\theta_1}\sin\varphi & -\dfrac{iY_S}{\sqrt{\dfrac{1}{\rho_1\kappa_1}}\cos\theta_1}\sin\varphi + \cos\varphi \end{bmatrix}, \quad (S4)$$

Set

$$M = \begin{bmatrix} A & B \\ C & D \end{bmatrix}, \tag{S5}$$

According to TMM, the transmission coefficient of the system is defined as

$$t = \frac{2}{A + BY_2 \cos\theta_2 + C/Y_2 \cos\theta_2 + D}. \tag{S6}$$

Here, $Y_2$, $\theta_2$, and $Y_2\cos\theta_2$ are the admittance, angle of incidence, and wave admittance of left/right medium, respectively. Hence, the denominator of Eq. (S6) can be expressed as

$$E = A + BY_2 \cos\theta_2 + C/Y_2 \cos\theta_2 + D$$

$$= 2\cos\varphi + \left( \frac{-iY_2 \cos\theta_2}{\sqrt{\frac{1}{\rho_1 \kappa_1}} \cos\theta_1} + \frac{-i\sqrt{\frac{1}{\rho_1 \kappa_1}} \cos\theta_1}{Y_2 \cos\theta_2} + \frac{i(Y_s)^2}{Y_2 \cos\theta_2 \sqrt{\frac{1}{\rho_1 \kappa_1}} \cos\theta_1} \right) \sin\varphi, \tag{S7}$$

According to the analysis in Part 2.1 of the main text, the system transmittance is $T = 1$ at EPs, and because $T = t^2$, the following expression holds:

$$|t| = \left| \frac{2}{2\cos\varphi + H\sin\varphi} \right| = 1. \tag{S8}$$

Here,

$$H = \frac{-iY_2 \cos\theta_2}{\sqrt{\frac{1}{\rho_1 \kappa_1}} \cos\theta_1} + \frac{-i\sqrt{\frac{1}{\rho_1 \kappa_1}} \cos\theta_1}{Y_2 \cos\theta_2} + \frac{i(Y_s)^2}{Y_2 \cos\theta_2 \sqrt{\frac{1}{\rho_1 \kappa_1}} \cos\theta_1}, \tag{S9}$$

We have

$$\left| \cos\varphi + \frac{H}{2} \sin\varphi \right| = 1, \tag{S10}$$

Hence, the above condition can be satisfied, when $H/2 = \pm i$. For the convenience of derivation, we set $H = h_1 + h_2 + h_3$, using Snell's law for simplification. The density and bulk modulus of the left/right mediums are $\rho_2$ and $\kappa_2$, respectively, which yields

$$h_1 = \frac{-iY_2 \cos\theta_2}{\sqrt{\frac{1}{\rho_1 \kappa_1} \cos\theta_1}}$$

$$= \frac{\rho_1 Y_2 \cos^2\theta_2}{i\cos\theta_2 \sqrt{\frac{\rho_1}{\kappa_1} - \frac{\rho_2 \sin^2\theta_2}{\kappa_2}}}, \tag{S11}$$

$$h_2 = \frac{-i\sqrt{\frac{1}{\rho_1\kappa_1}}\cos\theta_1}{Y_2 \cos\theta_2}$$

$$= \frac{\frac{1}{Y_2\kappa_1}}{i\cos\theta_2\sqrt{\frac{\rho_1}{\kappa_1} - \frac{\rho_2\sin^2\theta_2}{\kappa_2}}} - \frac{\frac{1}{Y_2}\rho_1^{-1}\frac{\rho_2}{\kappa_2}\sin^2\theta_2}{i\cos\theta_2\sqrt{\frac{\rho_1}{\kappa_1} - \frac{\rho_2\sin^2\theta_2}{\kappa_2}}}, \tag{S12}$$

$$h_3 = \frac{i(Y_S)^2}{Y_2\cos\theta_2\sqrt{\frac{1}{\rho_1\kappa_1}}\cos\theta_1}$$

$$= -\frac{(Y_S)^2 \frac{1}{Y_2}\rho_1}{i\cos\theta_2\sqrt{\frac{\rho_1}{\kappa_1} - \frac{\rho_2\sin^2\theta_2}{\kappa_2}}}. \tag{S13}$$

Therefore

$$H = \frac{\rho_1 Y_2 \cos^2\theta_2 + \frac{1}{Y_2\kappa_1} - \frac{1}{Y_2}\rho_1^{-1}\frac{\rho_2}{\kappa_2}\sin^2\theta_2 - (Y_S)^2\frac{1}{Y_2}\rho_1}{i\cos\theta_2\sqrt{\frac{\rho_1}{\kappa_1} - \frac{\rho_2\sin^2\theta_2}{\kappa_2}}}, \tag{S14}$$

$$= \pm 2i$$

The admittance equations of metasurfaces at EPs are obtained by further simplification:

$$Y_S = Y_2 \cos\theta_2 \pm \sqrt{\frac{1}{\rho_1\kappa_1} - \rho_1^{-2}\frac{\rho_2 \sin^2\theta_2}{\kappa_2}}, \tag{S15}$$

Again, using Snell's law for Eq. (S15), we obtain

$$\begin{aligned}Y_S &= Y_2\cos\theta_2 \pm \sqrt{\frac{1}{\rho_1\kappa_1} - \rho_1^{-2}\frac{\rho_2\sin^2\theta_2}{\kappa_2}} \\ &= Y_2\cos\theta_2 \pm \frac{1}{\rho_1\kappa_1}\sqrt{1-\sin^2\theta_1} \\ &= Y_2\cos\theta_2 \pm Y_1\cos\theta_1\end{aligned} \qquad (S16)$$

Therefore, according to Eq. (S16), there are two solutions for the left-side metasurface at the EPs, which correspond to the left-side metasurfaces of the CPA-amplifier and ARCL-ARCR systems, respectively (i.e., they correspond to Eqs. (1) and (3) in the main text, respectively). In addition, because the left-side and right-side metasurfaces satisfy the gain-loss balance, and their admittances are opposite to each other, the right-side metasurfaces at Eps, which were obtained by TMM, correspond to the right-side metasurfaces of the CPA-amplifier and ARCL-ARCR systems, respectively (i.e., they correspond to Eqs. (2) and (4) in the main text, respectively). Therefore, the conclusions that were derived from TMM are consistent with those obtained by the theoretical model, which verifies the accuracy of the theoretical model. In addition, the solutions of the EPs, which were obtained directly using the TMM, cannot accurately reflect the different physical mechanisms of the CPA-amplifier and ARCL-ARCR systems, and it also cannot accurately reflect the physical mechanisms of different metasurfaces. However, our theoretical model not only has a simpler form but also can reveal the physical mechanisms, and it is more scientifically significant.

**III. Simulated verification of the two inherent prerequisites needed for the existence of EPs**

For PT-symmetric acoustic metasurfaces, the inherent prerequisite for the existence of EPs is usually that the condition for total reflection cannot be satisfied, i.e., $\sqrt{\kappa_1\rho_2/\kappa_2\rho_1}\cdot\sin\theta_2 < 1$ (obtained from Snell's refraction law). This is, in this Letter referred to as the first inherent prerequisite. Moreover, previous research indicates that [SR8], if the admittance of a medium in a system is much smaller than that of other mediums, this medium tends to be neglected. Therefore, according to Eqs. (1)-(4) in the main text, if the difference between the wave admittances of middle and left/right mediums in the system is too large, the inability of the medium with the small wave admittance contributes to the realization of EPs. This makes it impossible to realize EPs. Thus, the second inherent prerequisite can be analyzed using our theoretical model. In other words, the difference between the wave admittances of middle and the left/right mediums cannot exceed one order of magnitude, i.e., $0.1 < \sqrt{\kappa_1 \rho_1 \cos^2\theta_2 / \kappa_2 \rho_2 \cos^2\theta_1} < 10$. The EPs of the PT-symmetric acoustic metasurfaces can exist only if both inherent prerequisites are satisfied simultaneously, even using our theoretical model. In acoustics, the difference between the wave admittances of different states of matter is quite large and has a significant impact on the realization of EPs. In this Letter, the system is simulated to verify the accuracy of the two inherent prerequisites - see Fig. S3.

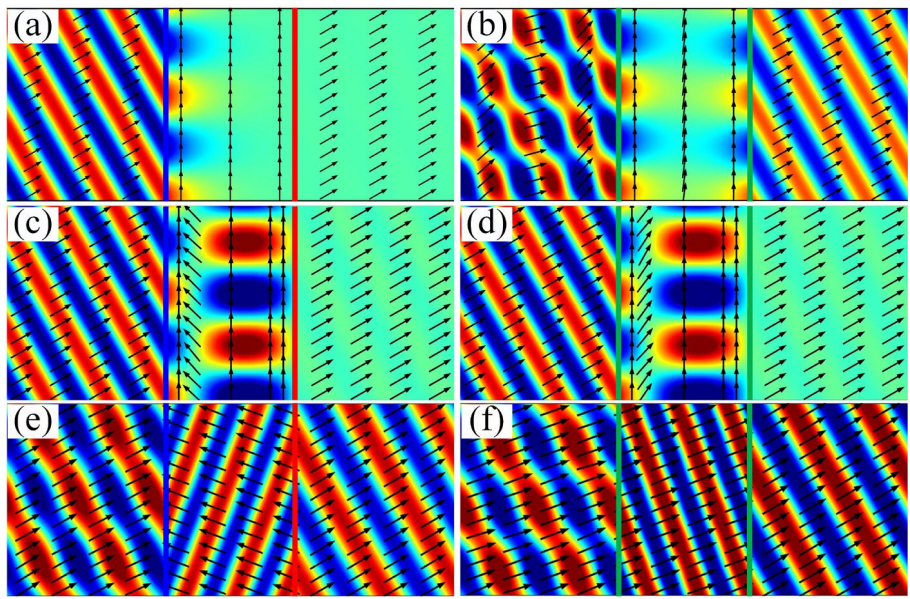

FIG. S3. Total sound pressure fields of the system, when the inherent prerequisites cannot be satisfied: (a) CPA-amplifier system, (b) ARCL-ARCR system, where the first inherent prerequisite cannot be satisfied, $\kappa_i = b_i\kappa_0$, $\rho_i = c_i\rho_0$, $\kappa_0=1.42\times10^5$ Pa, $\rho_0=1.2$ kg/m$^3$, $\lambda$=5mm, $b_1$=8, $c_1$=2, $b_2$=1.47, $c_2$=2.34, $\theta_2$=30°, $\sqrt{\kappa_1\rho_2/\kappa_2\rho_1}\cdot\sin\theta_2 \approx 1.26 > 1$. (c) CPA-amplifier system, (d) ARCL-ARCR system, where the second inherent prerequisite cannot be satisfied and the surrounding mediums belong to different states of mater, $\kappa_1 = b_1\kappa_{00}$, $\rho_1 = c_1\rho_{00}$, $\kappa_2 = b_2\kappa_{11}$, $\rho_2 = c_2\rho_{11}$, $\kappa_{00}$=2.18×10$^9$ Pa, $\rho_{00}$=1000 kg/m$^3$, $\kappa_{11}$=1.42×10$^5$ Pa, $\rho_{11}$=1.2 kg/m$^3$, $\lambda$=5mm, $b_1$=0.41, $c_1$=3.74, $b_2$=1.47, $c_2$=2.34, $\theta_2$=30°,

$Y_{W1}$=2.42×10$^{-7}$ m$^3$/(N×s), $Y_{W2}$= 1.13×10$^{-3}$ m$^3$/(N×s). (e) CPA-amplifier system, (f) ARCL-ARCR system, where the second inherent prerequisite cannot be satisfied and the surrounding mediums belong to the same state of mater, $\kappa_i = b_i\kappa_0$, $\rho_i = c_i\rho_0$, $\kappa_0$=1.42×10$^5$ Pa, $\rho_0$=1.2 kg/m$^3$, $\lambda$=5mm, $b_1$=0.141, $c_1$=0.274, $b_2$=5.47, $c_2$=5.34, $\theta_2$=30°, $Y_{W1}$= 1.15×10$^{-2}$ m$^3$/(N×s), $Y_{W2}$= 3.88×10$^{-4}$ m$^3$/(N×s). In all of the above cases, the parameters of metasurfaces are obtained using the theoretical model. $\kappa_i$, $\rho_i$, and $Y_{Wi}$ are the bulk modulus, density, and wave admittances of the surrounding mediums, respectively ($i$ = 1, 2, corresponds to middle or left/right medium of the PT-symmetric acoustic metasurfaces, respectively).

As can be seen from Figs. S3(a) and S3(b), the CPA-amplifier and ARCL-ARCR systems cannot be realized by the PT-symmetric metasurfaces (even using the theoretical model), when the first inherent prerequisite cannot be satisfied. When the surrounding mediums belong to different states of matter, it often causes the difference of their wave admittances to be greater than one order of magnitude, which cannot meet the second inherent prerequisite, and, thus, no EPs can exist - see Figs. S3(c) and S3(d). Moreover, even when the surrounding mediums belong to the same state of matter, there are cases where the difference in their wave admittances exceeds one order of magnitude, which also does not permit EPs - see Figs. S3(e) and S3(f).

Based on the above simulation, it can be concluded that, unlike the first inherent prerequisite for the EP's existence, which was reported in a previous study [SR9], the second inherent prerequisite *can* be analyzed using our theoretical model. Only if both inherent prerequisites are satisfied, however, can the EPs exist (i.e., the novel effects can be realized). These results confirm the inherent prerequisites for EPs in PT-symmetric acoustic metasurfaces.

## IV. Realizations of a CPA-amplifier system for different conditions

To consider PT-symmetric acoustic metasurfaces in gaseous systems, we set $\kappa_i = b_i\kappa_0$, $\rho_i = c_i\rho_0$. Here, $\kappa_0$ and $\rho_0$ are the bulk modulus and density of air, respectively, while $b_i$ and $c_i$ are arbitrary constants ($i$=1, 2, corresponds to the middle or left/right mediums of the PT-symmetric acoustic metasurfaces, respectively). All of the above parameters satisfy the two inherent prerequisites for EPs. If the middle and left/right mediums and angle of incidence are determined, the wave admittances of the middle and left/right mediums can be calculated using the Eqs. (5) and (6) in the main text. Then, the admittances of metasurfaces, which required the realization of a CPA-amplifier system, can be obtained by substituting the wave admittances in Eqs. (1) and (2) in the main text. Furthermore, the corresponding complex bulk modulus of the metasurfaces can be obtained using Eq. (7) in the main text. Together with the metasurface thickness and the incident wavelength, the corresponding total acoustic pressure fields are shown in Fig. S4.

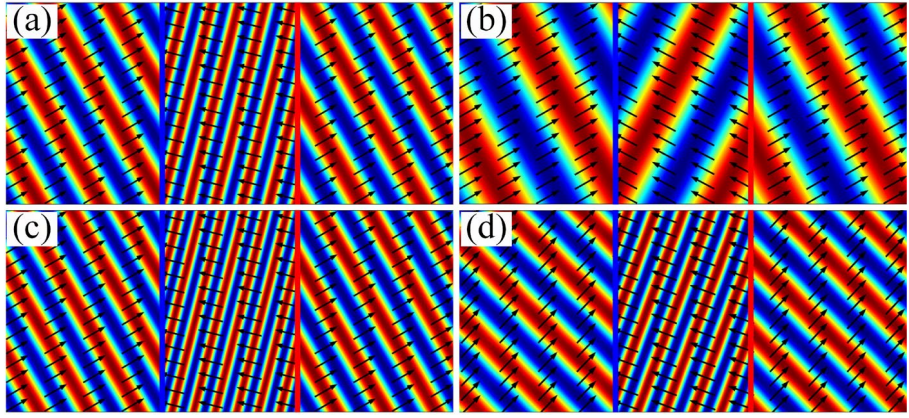

FIG. S4. The total sound pressure fields of the CPA-amplifier system for different surrounding mediums, different wavelengths, and different angles of incidence: (a) $\lambda=5$mm, $b_1=0.41$, $c_1=2.74$, $b_2=1.47$, $c_2=2.34$, $\theta_2=30°$, $Y_{W1}=2.22\times10^{-3}$ m$^3$/(N×s), $Y_{W2}=1.13\times10^{-3}$ m$^3$/(N×s). (b) $\lambda=5$mm, $b_1=2.41$, $c_1=1.21$, $b_2=3.47$, $c_2=1.54$, $\theta_2=30°$, $Y_{W1}=1.25\times10^{-3}$ m$^3$/(N×s), $Y_{W2}=9.08\times10^{-4}$ m$^3$/(N×s). (c) $\lambda=4$mm, $b_1=0.41$, $c_1=2.74$, $b_2=1.47$, $c_2=2.34$, $\theta_2=30°$, the wave admittances of the surrounding mediums are the same as in (a). (d) $\lambda=5$mm, $b_1=0.41$, $c_1=2.74$, $b_2=1.47$, $c_2=2.34$, $\theta_2=45°$, $Y_{W1}=2.15\times10^{-3}$ m$^3$/(N×s), $Y_{W2}=9.24\times10^{-4}$ m$^3$/(N×s). The parameters of the metasurfaces are obtained using the theoretical model, and $Y_{W1}$ and $Y_{W2}$ are the wave admittances of the middle and left/right mediums, respectively.

As can be seen from Fig. S4, the parameters of the metasurfaces, which are needed to realize the CPA-amplifier system for different surrounding mediums, different wavelengths, and different angles of incidence, can be obtained using the theoretical model – provided that the two inherent prerequisites are satisfied.

To further compare the unidirectional antireflection negative refraction and ordinary negative refraction, the ordinary negative refraction of the negative refractive material is analyzed in the absence of PT-symmetric acoustic metasurfaces. All other parameters are consistent with Fig. S4, whose total sound-pressure fields are shown in Fig. S5.

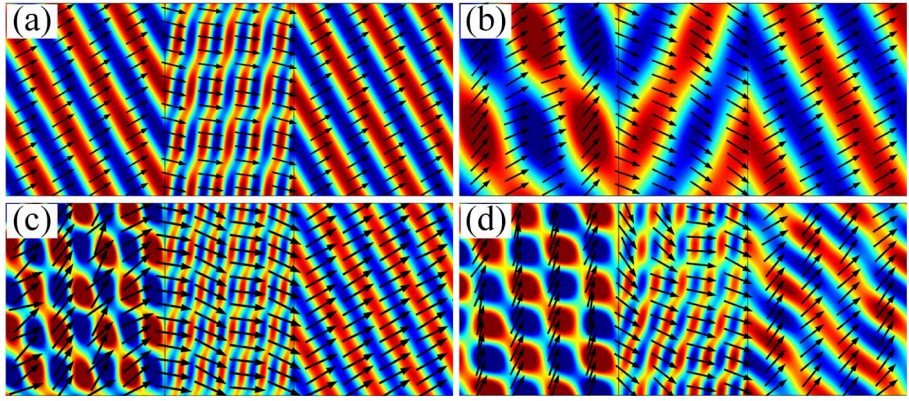

FIG. S5. Total sound pressure fields of the ordinary negative refraction for different mediums, different wavelengths, and different angles of incidence: (a) $\lambda=5$mm, $b_1=-0.41$, $c_1=-2.74$, $b_2=1.47$, $c_2=2.34$, $\theta_2=30°$. (b) $\lambda=5$mm, $b_1=-2.41$, $c_1=-1.21$, $b_2=3.47$, $c_2=1.54$, $\theta_2=30°$. (c) $\lambda=4$mm, $b_1=-0.41$, $c_1=-2.74$, $b_2=1.47$, $c_2=2.34$, $\theta_2=30°$. (d) $\lambda=5$mm, $b_1=-0.41$, $c_1=-2.74$, $b_2=1.47$, $c_2=2.34$, $\theta_2=45°$.

As can be seen from Fig. S5, the impedance of the negative refractive material (middle medium) for different conditions does not match the impedance of the left/right medium. Furthermore, there are reflections in the system, which reduce the transmittance of the system. At the same time, the reflected acoustic waves interfere with the incident acoustic waves, which destroys their waveform.

## V. Realization of an ARCL-ARCR system for different conditions

To verify the accuracy of the theoretical model of the ARCL-ARCR system, the required metasurface parameters (to realize the ARCL-ARCR system for different conditions) can also be obtained using the theoretical model (the method is the same as described in Part IV). The surrounding mediums are the same as in Fig. S4, the total sound pressure fields of the ARCL-ARCR system are as shown in Fig. S6.

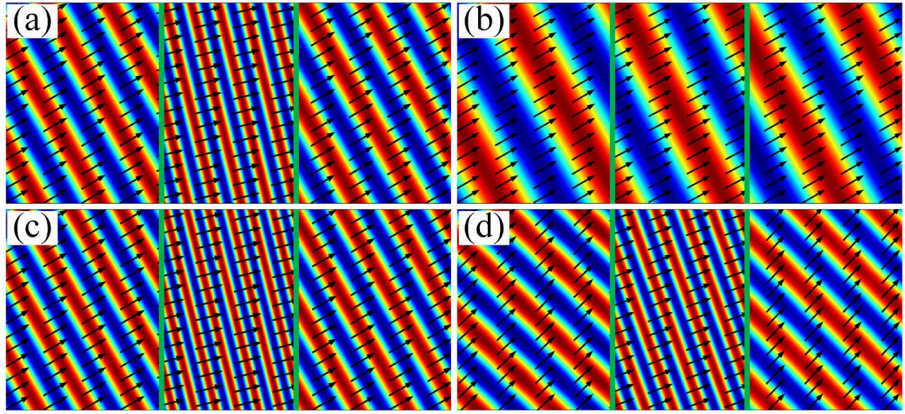

FIG. S6. The total sound pressure fields of the ARCL-ARCR system for different surrounding mediums, different wavelengths, and different angles of incidence: (a) $\lambda$=5mm, $b_1$=0.41, $c_1$=2.74, $b_2$=1.47, $c_2$=2.34, $\theta_2$=30°. (b) $\lambda$=5mm, $b_1$=2.41, $c_1$=1.21, $b_2$=3.47, $c_2$=1.54, $\theta_2$=30°. (c) $\lambda$=4mm, $b_1$=0.41, $c_1$=2.74, $b_2$=1.47, $c_2$=2.34, $\theta_2$=30°. (d) $\lambda$=5mm, $b_1$=0.41, $c_1$=2.74, $b_2$=1.47, $c_2$=2.34, $\theta_2$=45°. The parameters of the metasurfaces are obtained using the theoretical model, and the wave admittances of the surrounding mediums correspond to those in Fig. S4 in the same order.

From Fig. S6, it can be seen that the ARCL-ARCR system can also be realized using the theoretical model for different surrounding mediums, different wavelengths, and different angles of incidence, provided the two inherent prerequisites are satisfied. To further compare the unidirectional antireflection positive refraction with ordinary positive refraction, the ordinary positive refraction is analyzed in the absence of the PT-symmetric acoustic metasurfaces. The total sound pressure fields are shown in Fig. S7.

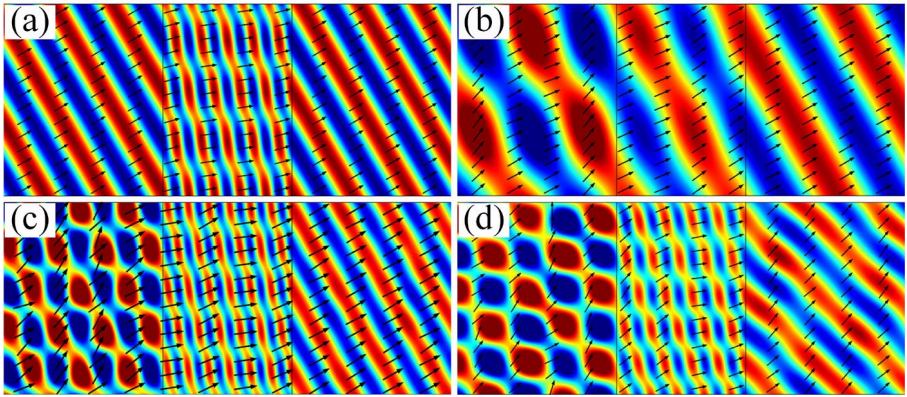

FIG. S7. Total sound pressure fields of the ordinary positive refraction, for different mediums, different wavelengths, and different angles of incidence: (a) $\lambda$=5mm, $b_1$=-0.41, $c_1$=-2.74, $b_2$=1.47,

$c_2$=2.34, $\theta_2$=30°. (b) $\lambda$=5mm, $b_1$=-2.41, $c_1$=-1.21, $b_2$=3.47, $c_2$=1.54, $\theta_2$=30°. (c) $\lambda$=4mm, $b_1$=-0.41, $c_1$=-2.74, $b_2$=1.47, $c_2$=2.34, $\theta_2$=30°. (d) $\lambda$=5mm, $b_1$=-0.41, $c_1$=-2.74, $b_2$=1.47, $c_2$=2.34, $\theta_2$=45°.

A comparison between Figs. S6 and S7 shows that unidirectional antireflection positive refraction can be achieved by the PT-symmetric acoustic metasurfaces, which correspond to the ARCL-ARCR system. On the other hand, when the PT-symmetric acoustic metasurfaces are not present, the impedance mismatch leads to a low transmittance a strong reflection, which is not favorable for practical applications.

**VI. A simple theoretical model for arbitrary left/right mediums**

In this section, we further confirm that the theoretical model can be applied to arbitrary left/right mediums. We used the same method as shown in Fig. 3 in the main text to analyze arbitrary left/right mediums of the system. The results are shown in Fig. S8.

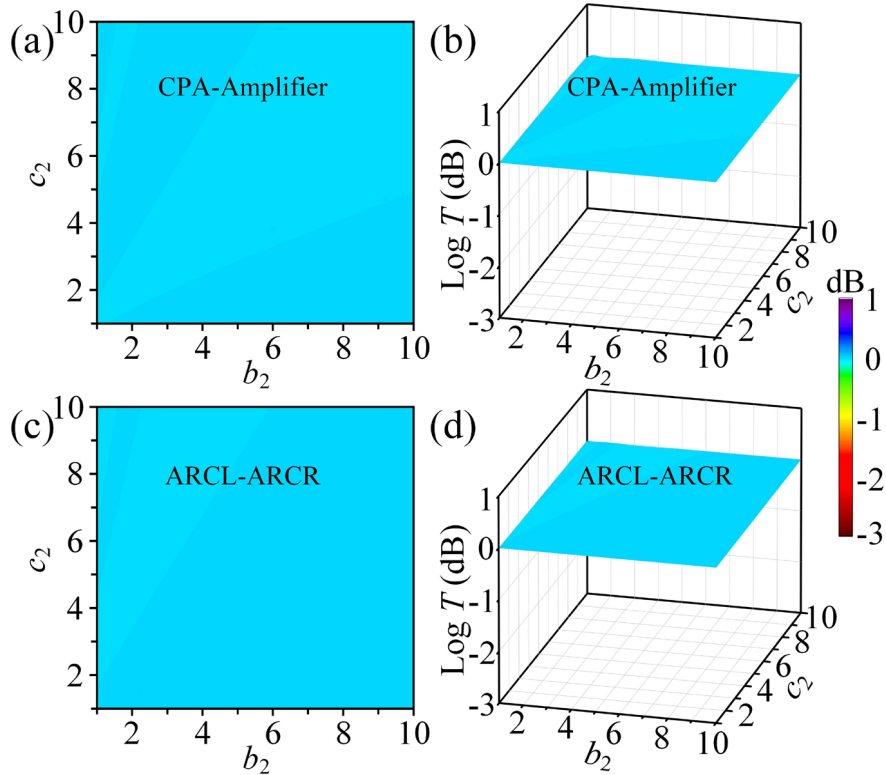

FIG. S8. Log$T$ of the system calculated by sweeping $b_1$ and $c_1$ in the range of [1,10]. Here, the parameters of the metasurfaces are obtained with the theoretical model and $\lambda$=5mm, $b_1$=0.41, $c_1$=2.74, $\theta_2$=30°: (a) CPA-amplifier system. (b) Three-dimensional view of (a). (c) ARCL-ARCR

system. (d) Three-dimensional view of (b).

As can be seen from Fig. S8, in the range [1,10], the Log$T$ of two systems is 0 dB for arbitrary left/right mediums. Hence, total transmission can be achieved for the two systems. This means that the required metasurface parameters to realize the CPA-amplifier and ARCL-ARCR systems can be obtained using the theoretical model for arbitrary left/right mediums.

**VII. Simple theoretical model for arbitrary incident wavelengths and angles**

To verify that the theoretical model is also universally applicable to arbitrary wavelengths and angles of incidence, the wavelengths and angles of incidence are also analyzed separately, using the same method as shown in Fig. S8. All other parameters remained fixed to the values used in Fig. S4(a). The results are shown in Fig. S9.

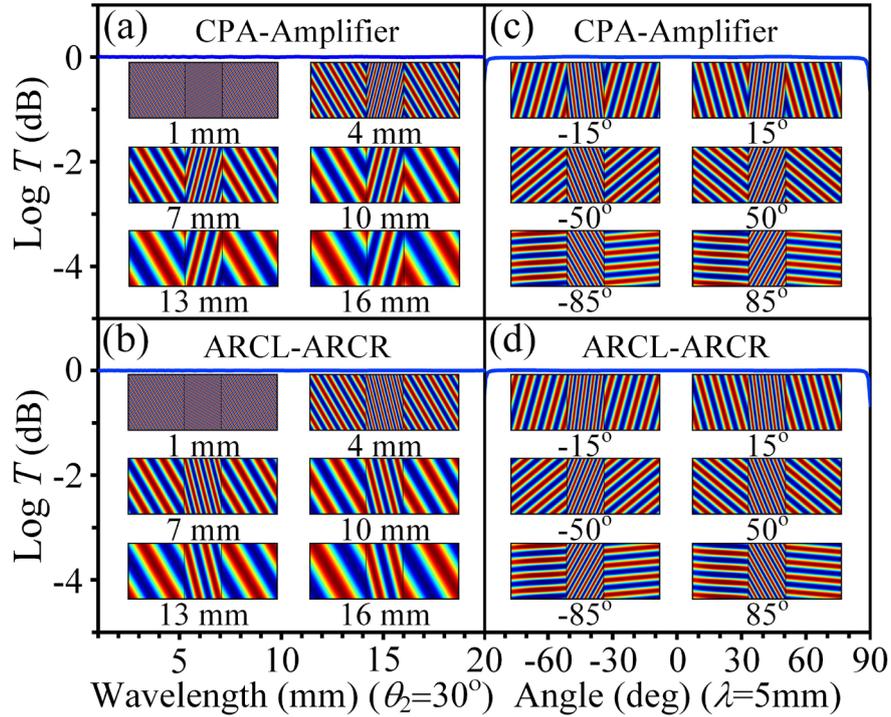

FIG. S9. Log$T$ of the system, which is calculated by sweeping the incident wavelengths $\lambda$ in the range [1 mm, 20 mm]: (a) CPA-amplifier system, (b) ARCL-ARCR systems. Log$T$ of the system, which is calculated using sweeping angles of incidence $\theta_2$ in the range [-90°, 90°]: (c) CPA-amplifier system, (d) ARCL-ARCR system. In all of the above cases, the insets show the corresponding total sound pressure fields, and the parameters of metasurfaces are obtained by the theoretical model with $b_1$=0.41, $c_1$=2.74, $b_2$=1.47, $c_2$=2.34.

From Fig. S9, it can be seen that the theoretical model is also universally applicable to incident wavelengths and angles. Compared to previous studies [SR9-SR11], the theoretical model is more general, has a simpler form, and is more helpful to distinguish the EPs of different systems based on their physical mechanisms.

## VIII. The simple theoretical model of gain-loss-imbalanced PT-symmetric acoustic metasurfaces

With our theoretical model, the metasurface parameters, which are required to realize the CPA-amplifier and ARCL-ARCR systems, can even be derived when the middle, left, and right mediums are different ordinary mediums. In addition, they do not need to satisfy a specific gain-loss ratio. In these instances, the systems consist of gain-loss-imbalanced PT-symmetric metasurfaces, and, by using the same method as in Fig. S1 (to analyze these systems), the corresponding models for gain-loss-imbalanced PT-symmetric acoustic metasurfaces can be obtained - see Fig. S10.

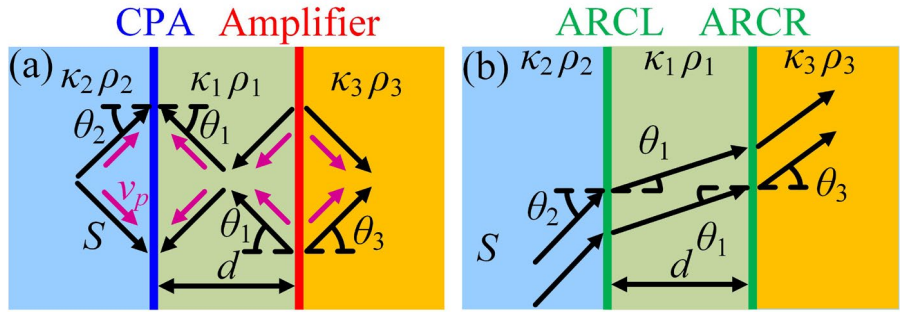

FIG. S10. Gain-loss-imbalanced PT-symmetric acoustic metasurfaces systems: (a) CPA-amplifier system, (b) ARCL-ARCR system.

We set $\kappa_i = b_i \kappa_0$, $\rho_i = c_i \rho_0$, where $\kappa_0$ and $\rho_0$ denote the bulk modulus and density of air, and $b_i$ and $c_i$ are the arbitrary constants that ensure two inherent prerequisites are satisfied ($i$=1, 2, 3, corresponds to the middle, left, or right medium of the

PT-symmetric acoustic metasurfaces, respectively). Eqs. (1)-(4) in the main text are transformed into the equations applied to the model shown in Fig. S10 using the theoretical model (the modeling idea and the derivation of the equation can be found in Fig. 1 of the main text) as follows:

$$Y_{CPA} = Y_1 \cos\theta_1 + Y_2 \cos\theta_2, \tag{S17}$$

$$Y_{Amplifier} = -Y_1 \cos\theta_1 - Y_3 \cos\theta_3, \tag{S18}$$

$$Y_{ARCL} = -Y_1 \cos\theta_1 + Y_2 \cos\theta_2, \tag{S19}$$

$$Y_{ARCR} = Y_1 \cos\theta_1 - Y_3 \cos\theta_3. \tag{S20}$$

Using the above equations, it becomes clear that, if the admittances of the three mediums are not equal, the admittances that correspond to gain and loss are no longer opposite to each other. This scenario is different from the conventional PT-symmetric system with gain-loss balance.

According to Fig. S10 and Eqs. (S17)-(S20), the CPA-amplifier and ARCL-ARCR systems with gain-loss imbalance can be obtained when $d_S$ and $d$ are the same as in Fig. S4. Their total sound pressure fields are shown in Fig. S11.

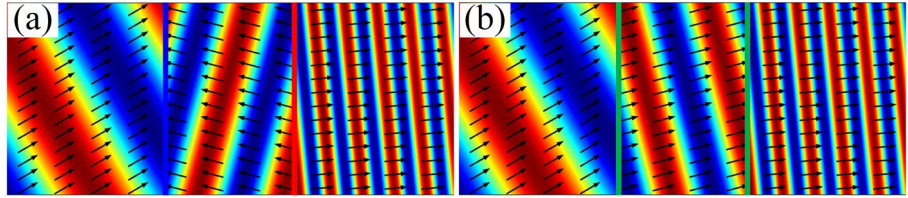

FIG. S11. Total sound pressure fields of the CPA-amplifier and ARCL-ARCR systems for the gain-loss-imbalanced system: (a) CPA-amplifier system, (b) ARCL-ARCR system, where the parameters of metasurfaces are obtained using the theoretical model, $\lambda$=5mm, $b_1$=2.37, $c_1$=1.43, $b_2$=1.35, $c_2$=0.21, $b_3$=2.12, $c_3$=8.73, $\theta_2$=30°, $Y_{W1}$= 1.28×10$^{-3}$ m$^3$/(N×s), $Y_{W2}$=3.94×10$^{-3}$ m$^3$/(N×s), $Y_{W3}$= 5.60×10$^{-4}$ m$^3$/(N×s). $Y_{Wi}$ denotes the wave admittance of the surrounding mediums ($i$=1, 2, 3, corresponds to the middle, left, or right medium, respectively).

It can be seen from Fig. S11, if the middle, left and right mediums are different ordinary mediums, i.e., their wave admittances are not equal, and both the CPA-amplifier and ARCL-ARCR systems can still be obtained with the theoretical model. Hence, the PT-symmetric acoustic metasurfaces do not need to meet the conditions for the gain-loss-balanced system as in previous studies, which makes them more universally applicable and valuable.

To further verify that the theoretical model can still be universally applicable when gain and loss satisfy arbitrary ratios, different gain-loss ratios are analyzed. We set the ratio $r = |Y_{SL}/Y_{SR}|$, where $Y_{SL}$ and $Y_{SR}$ are the admittances of the left-side and right-side metasurfaces in this system, respectively. The admittances can be calculated using Eqs. (S17)-(S20). Furthermore, to make the analytical results more reliable, we sweep $r$ in the range of [1, 10]. This is done because, according to Eqs. (S17)-(S20), the limit of $r$ is 10 for the two inherent preconditions, and the variation of $r$ reaches an order of magnitude, which can fully explain the stability of the analytical results and the universality of the research conclusions. The obtained Log$T$ and total sound pressure fields are shown in Fig. S12.

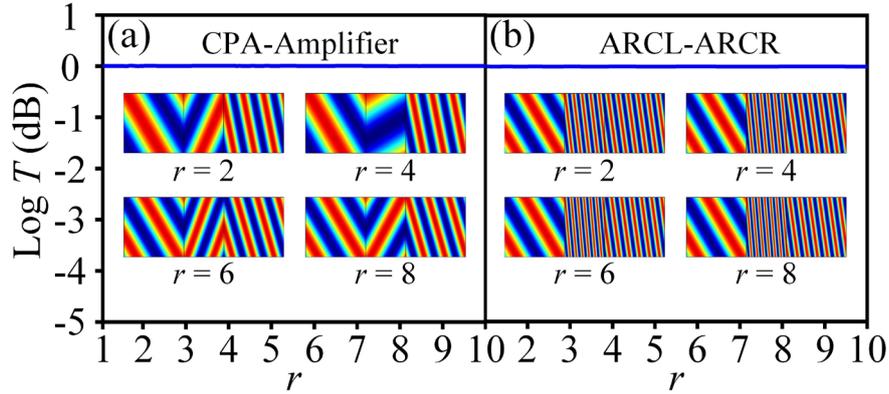

FIG. S12. Log$T$ of the system, which is calculated by sweeping $r$ in the range [1, 10]: (a) CPA-amplifier system, $b_i$ and $c_i$ are adjusted to make $r$ assume different values in the range, where $i = 1, 2, 3$. (b) ARCL-ARCR system, $b_2$=2, $c_2$=0.6, $b_3$=2.12, $c_3$=8.73, $b_1$, and $c_1$ are adjusted to ensure $r$ assumes different values across the sweeping range. In all of the above cases, the insets show the corresponding total sound pressure fields, and the parameters of the metasurfaces are obtained using the theoretical model.

As can be seen from Fig. S12, the theoretical model is applicable when both gain and loss satisfy arbitrary ratios and the two inherent prerequisites are met.

## IX. Realization of EPs when the complex parameters of metasurfaces are fixed

The theoretical model can also be used to obtain the required surrounding ordinary mediums parameters of EPs via inverse derivation when the complex parameters of the metasurfaces are fixed. If the complex bulk modulus on the left-side and metasurfaces on the right-side are $\kappa_{SL}$ and $\kappa_{SR}$, respectively, and all other conditions are determined, the required relationships between the middle, left, and right mediums can be obtained using Eqs. (S17)-(S20). In addition, the middle, left, and right mediums also need to satisfy Snell's law:

$$\sqrt{\rho_1/\kappa_1} \cdot \sin\theta_1 = \sqrt{\rho_2/\kappa_2} \cdot \sin\theta_2 = \sqrt{\rho_3/\kappa_3} \cdot \sin\theta_3 \qquad (S21)$$

For PT-symmetric acoustic metasurfaces with fixed complex parameters, according to the above analysis, if the angle of incidence, incident wavelength, and thickness of metasurface are fixed first, then one or two regional mediums are fixed and the parameters of the medium in the other region can be obtained using the theoretical model. However, if the gain and loss are *balanced*, only one regional medium needs to be fixed. If the gain and loss are *imbalanced*, two regional mediums need to be fixed. Because gain-loss-unbalanced PT-symmetric acoustic metasurfaces are taken into account, the system is constructed using the model shown in Fig. S10. It corresponds to a gain-loss-balanced system, provided $b_2=b_3$ and $c_2=c_3$. If $b_2 \neq b_3$ or $c_2 \neq c_3$, it corresponds to gain-loss-unbalanced system, and the total sound pressure fields in both cases are as shown in Fig. S13.

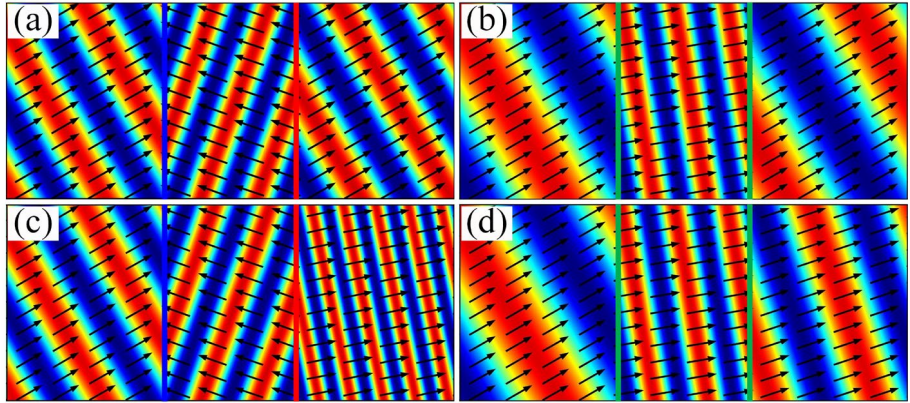

FIG. S13. Total acoustic-pressure fields for the CPA-amplifier and ARCL-ARCR systems, when the complex parameters of the metasurfaces are fixed. The required surrounding ordinary medium

parameters are obtained using the theoretical model, while all other parameters are fixed ($\theta_2=30°$, $\lambda=5$ mm, $d_S=0.001\lambda$). Shown is the gain-loss-*balanced* system: (a) CPA-amplifier system, (b) ARCL-ARCR system, where the complex parameters for the metasurfaces and middle medium are fixed ($\kappa_{SL}=0.044i\times\kappa_0$, $\kappa_{SR}=-0.044i\times\kappa_0$, $b_1=19$, $c_1=30$). Shown is the gain-loss-*unbalanced* system: (c) CPA-amplifier system, where the parameters of the left medium are $b_2=10$, $c_2=7$, (d) ARCL-ARCR system, where the parameters of the left medium are $b_2=11$, $c_2=2$, and the parameters of the metasurfaces, middle medium are fixed for the two above cases ($\kappa_{SL}=0.044i\times\kappa_0$, $\kappa_{SR}=-0.073i\times\kappa_0$, $b_1=19$, $c_1=30$).

Figs. S13(a) and S13(b) show that for gain-loss-balanced PT-symmetric acoustic metasurfaces, when the parameters of PT-symmetric acoustic metasurfaces and middle medium are fixed, the left/right medium, which are required to realize the CPA-amplifier and ARCL-ARCR systems, can be deduced using the theoretical model. Hence, this method can effectively reduce the difficulty of implementing PT-symmetric acoustic metasurfaces. Furthermore, for gain-loss-imbalanced PT-symmetric acoustic metasurfaces (when the parameters of PT-symmetric acoustic metasurfaces, middle and left mediums are fixed), the right medium that is required to realize the CPA-amplifier and ARCL-ARCR systems, can be deduced by the model. Overall, this method requires less gain, which has many practical benefits.

Based on the above results, clearly, when the complex parameters of PT-symmetric acoustic metasurfaces are determined, the real parameters of surrounding mediums, which are needed to create EPs, can be derived using the theoretical model. This makes it possible to convert the difficult complex-parameter problem into an easy-to-implement real-parameter problem and provide theoretical guidance to develop a real-life PT-symmetric system.

## X. Realization of a CPA-amplifier and ARCL-ARCR systems for different states of matter

To further demonstrate that our theoretical model can be used for arbitrary states of matter, $\kappa_0$ and $\rho_0$ are set to the bulk modulus, and density of water, and steel, respectively. The total acoustic pressure fields are shown in Fig. S14.

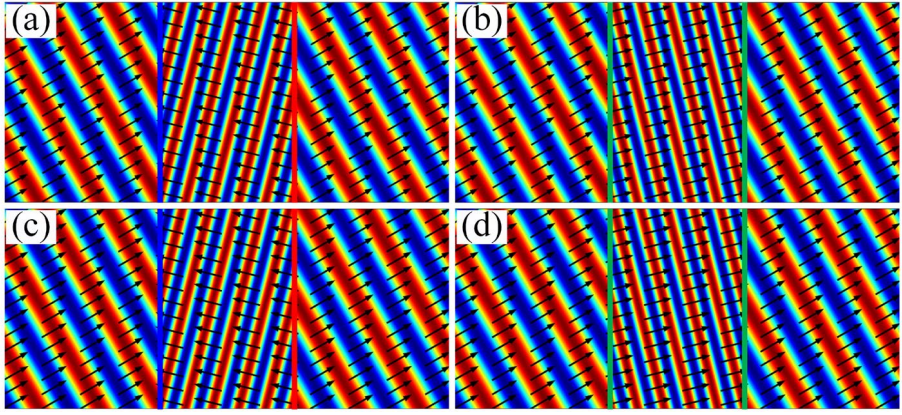

FIG. S14. The total sound pressure fields for the CPA-amplifier and ARCL-ARCR systems for different states of matter, where the parameters of metasurfaces are obtained using the theoretical model: (a)(b) The liquid state, with $\kappa_0=2.18\times10^9$ Pa, $\rho_0=1000$ kg/m$^3$, $\lambda=5$mm, $b_1=0.41$, $c_1=2.74$, $b_2=1.47$, $c_2=2.34$, $\theta_2=30^o$, $Y_{W1}=3.16\times10^{-7}$ m$^3$/(N×s), $Y_{W2}=6.20\times10^{-7}$ m$^3$/(N×s). (c)(d) The solid state, with $\kappa_0=2.08\times10^{11}$ Pa, $\rho_0=7850$ kg/m$^3$, $\lambda=5$mm, $b_1=0.41$, $c_1=2.74$, $b_2=1.47$, $c_2=2.34$, $\theta_2=30^o$, $Y_{W1}=2.26\times10^{-8}$ m$^3$/(N×s), $Y_{W2}=1.16\times10^{-8}$ m$^3$/(N×s). $Y_{W1}$ and $Y_{W2}$ are the wave admittances of the middle and left/right mediums, respectively.

From Fig. S14, it can be seen that CPA-amplifier and ARCL-ARCR systems can also be obtained using the theoretical model for different states of matter. This suggests the general applicability of the theoretical model and its practical application value.

## XI. Construction guidelines for real-life acoustic PT-symmetric systems

In previous studies, researchers have successfully constructed real-life acoustic PT-symmetric systems using active loudspeakers and absorbing structures [SR13-SR15]. These achievements provide valuable ideas for the construction of PT-symmetric acoustic metasurfaces in real-life. Because the realization of the above PT-symmetric real system necessarily involves acoustic gain, it is more difficult to obtain real-life PT-symmetric systems. The EPs, however, can be created with the theoretical model even when both gain and loss of the PT-symmetric metasurfaces satisfy arbitrary ratios. Therefore, according to the model in Fig. S10, Eqs. (S19), and

(S20), the PT-symmetric system, which consists of only two passive metasurfaces, can be constructed with the theoretical model. In this case, the wave admittances of the surrounding mediums need to meet the following relation: $\sqrt{\cos^2\theta_1/\kappa_1\rho_1} > \sqrt{\cos^2\theta_2/\kappa_2\rho_2} > \sqrt{\cos^2\theta_3/\kappa_3\rho_3}$. In other words, passive PT-symmetric metasurface systems can be constructed by tuning the parameters of the surrounding mediums. This reduces the difficulty to make real-life PT-symmetric systems, significantly. The results are shown in Fig. S15.

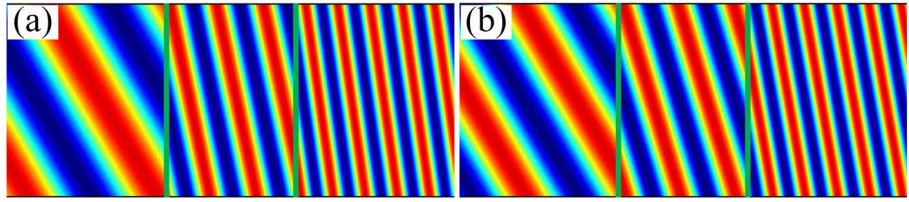

FIG. S15. Total sound pressure fields of the ARCL-ARCR system. The PT-symmetric acoustic metasurfaces are all passive, with fixed $\lambda$=5mm, $\theta_2$=30°: (a) $b_1$=2, $c_1$=4, $b_2$=3, $c_2$=1, $b_3$=2.12, $c_3$=8.73, $Y_{SL}$=3.73×10$^{-4}$ m$^3$/(N×s), $Y_{SR}$=1.40×10$^{-3}$ m$^3$/(N×s). (b) $b_1$=2, $c_1$=3, $b_2$=2, $c_2$=1, $b_3$=2.12, $c_3$=8.73, $Y_{SL}$=5.37×10$^{-4}$ m$^3$/(N×s), $Y_{SR}$=1.50×10$^{-3}$ m$^3$/(N×s).

In a real-life system, sufficient gain is much harder to realize, and the strict gain-loss-balance-requirement hinders the EPs realization of PT-symmetric acoustic metasurfaces [SR16]. However, these drawbacks can be overcome with the help of the theoretical model, which can help realize EPs at arbitrary gain-loss ratios - see Fig. [SR15]. In addition, in Part IX, the EPs are obtained with surrounding ordinary mediums, where the complex parameters of PT-symmetric acoustic metasurfaces are fixed. In other words, a real-life PT-symmetric system can, in fact, be accomplished based on the already existing constructed acoustic gain and loss. This has great significance for finding practical ways to obtain real PT-symmetric systems.

**References**

[SR1] W. J. Yu, S. H. Chae, S. Y. Lee, D. L. Duong, and Y. H. Lee, "Ultra-transparent, flexible single-walled carbon nanotube non-volatile memory device with an oxygen-decorated graphene electrode." Adv. Mater. 23(16):

1889-1893 (2011).

[SR2]  J. Luo, Y. Yang, Z. Yao, W. Lu, B. Hou, Z. H. Hang, C. T. Chan, and Y. Lai, "Ultratransparent media and transformation optics with shifted spatial dispersions," Phys. Rev. Lett. 117(22): 223901 (2016).

[SR3]  Z. Yao, J. Luo, and Y. Lai, "Photonic crystals with broadband, wide-angle, and polarization-insensitive transparency," Opt. Lett. 41(21): 5106-5109 (2016).

[SR4]  Z. Yao, J. Luo, and Y. Lai, "Illusion optics via one-dimensional ultratransparent photonic crystals with shifted spatial dispersions," Opt. Express 25(25): 30931-30938 (2017).

[SR5]  F. Bongard, H. Lissek, and J. R. Mosig, "Acoustic transmission line metamaterial with negative/zero/positive refractive index," Phys. Rev. B 82(9): 094306 (2010).

[SR6]  C. Shen, J. Xu, N. X. Fang, and Y. Jing, "Anisotropic complementary acoustic metamaterial for canceling out aberrating layers," Phys. Rev. X 4(4): 041033 (2014).

[SR7]  J. Zhang, Y. Cheng, and X. Liu, "Extraordinary acoustic transmission at low frequency by a tunable acoustic impedance metasurface based on coupled Mie resonators," Appl. Phys. Lett. 110(23): 233502 (2017).

[SR8]  C. K. Alexander and M. Sadiku, Fundamentals of electric circuit,6th ed. (McGraw-Hill Education, 2016).

[SR9]  J. Luo, J. Li, and Y. Lai, "Electromagnetic impurity-immunity induced by parity-time symmetry," Phys. Rev. X 8(3): 031035 (2018).

[SR10] R. Fleury, D. L. Sounas, and A. Alù, "Negative refraction and planar focusing based on parity-time symmetric metasurfaces," Phys. Rev. Lett. 113(2): 023903 (2014).

[SR11] F. Monticone, C. A. Valagiannopoulos, and A. Alù, "Parity-Time symmetric nonlocal metasurfaces: all-angle negative refraction and volumetric imaging," Phys. Rev. X 6(4): 041018 (2016).

[SR12] D. M. Pozar, Microwave Engineering, 3$^{rd}$ ed. (Wiley, New York, 2011).

[SR13] R. Fleury, D. Sounas, and A. Alù, "An invisible acoustic sensor based on

parity-time symmetry," Nat. Commun. 6(1): 5905 (2015).

[SR14] C. Shi, M. Dubois, Y. Chen, L. Cheng, H. Ramezani, Y. Wang, and X. Zhang, "Accessing the exceptional points of parity-time symmetric acoustics," Nat. Commun. 7(1): 11110 (2016).

[SR15] H. Li, M. Rosendo-López, Y. Zhu, X. Fan, D. Torrent, B. Liang, J. Cheng, and J. Christensen, "Ultrathin acoustic parity-time symmetric metasurface cloak," Research 2019: 8345683 (2019).

[SR16] H. Li, A. Mekawy, A. Krasnok, and A. Alù, "Virtual parity-time symmetry," Phys. Rev. Lett. 124(19): 193901 (2020).